\documentclass[12pt]{article}
\usepackage{amsmath}
\usepackage{epsfig}
\usepackage{psfrag}
\usepackage{amssymb}
\usepackage{graphicx}
\newtheorem{thm}{Theorem}

\newtheorem{lemma}{Lemma}
\newtheorem{propo}{Proposition}

\def\ex{\widehat{x}}

\def\E{{\mathbb E}}
\def\prob{{\mathbb P}}
\def\d{{\rm d}}
\def\ind{{\mathbb I}}

\def\R{{\mathbb R}}
\def\J{{\mathbb J}}

\def\lh{\widehat{\lambda}}
\def\gh{\widehat{\gamma}}
\def\Gh{\widehat{\Gamma}}

\def\us{\underline{s}}
\def\uy{\underline{y}}
\def\si{{\mathbb S}}

\def\cS{{\cal S}}
\def\Tr{{\rm Tr}}

\def\el{L}
\def\gee{G}

\def\geeh{\widehat{G}}
\def\dg{\delta\gamma}
\def\dgh{\delta\widehat{\gamma}}

\def\ux{\underline{x}}
\def\uy{\underline{y}}
\def\uw{\underline{w}}
\def\uo{\underline{\omega}}

\begin{document}
\renewcommand{\textfraction}{0}

\title{Belief Propagation Based Multi--User Detection}
\author{
\normalsize  Andrea Montanari \vspace*{3pt} \\
\small Laboratoire de Physique Th\'{e}orique\vspace*{-3pt} \\ 
\small Ecole Normale Sup\'{e}rieure\vspace*{-3pt} \\
\small 75005 Paris, FRANCE\vspace*{-3pt} \\
\small {\tt montanar@lpt.ens.fr}\and
\normalsize Balaji Prabhakar \vspace*{3pt} \\
\small Dept. of Electrical Engineering \vspace*{-3pt} \\  
\small Stanford University \vspace*{-3pt} \\
\small Stanford, CA 94305 \vspace*{-3pt} \\
\small {\tt balaji@stanford.edu} \and
\normalsize David Tse \vspace*{3pt} \\ 
\small  Dept. of Electrical Engineering and Computer Sciences\vspace*{-3pt} \\
\small  University of California, Berkeley \vspace*{-3pt} \\
\small {\tt  dtse@eecs.berkeley.edu }}
\date{}

\maketitle \thispagestyle{empty}

\parskip 8pt

\abstract{We apply belief propagation (BP) to multi--user detection
in a spread spectrum system, under the assumption of Gaussian symbols.
We prove that BP is both convergent and allows
to estimate the correct conditional expectation of the input symbols.
It is therefore an optimal --minimum mean square error-- detection algorithm.
This suggests the possibility of designing BP detection algorithms
for more general systems.

As a byproduct we rederive the Tse-Hanly formula for minimum mean square
error without any recourse to random matrix theory.}
%
%
\section{Introduction}
\label{sec:Intro}

Consider the multiuser detection problem of $K$ users, each spreading its symbol onto $N$ chips  in a spread spectrum system
\begin{equation}
\uy = \sum_{i=1}^K x_i \us_i + \uw\, .\label{eq:Model}
\end{equation}
Here $x_i\in\R$ is the symbol transmitted by user $i$, $s_i\in\R^N$ is 
the $N$-chip long signature sequence of user $i$, $y$ is the received vector, and $w$ is an $N\times 1$ vector of
i.i.d.\ mean 0 and variance $\sigma^2$ Gaussians.  We assume that the $x_i$
are i.i.d.\ with $\E\, x_i=0$ and $\E\, x_i^2 = 1$.
Finally, we shall denote by $\si$ the $N\times K$ signature matrix 
whose columns are the vectors $\us_i$.

Several years ago, Tse and Hanly \cite{TseHanly}, and Verd\'u and Shamai
\cite{VerduShamai} considered the case in which the 
symbols $x_i$'s are Gaussian random variables, which models the situation when they are ideal Shannon-coded symbols.   The signature sequences
were assumed themselves random, but known both at the receiver. More precisely
(here and in the following $A^\dagger$ denotes the transpose of
matrix, or vector, $A$)
\begin{eqnarray}
s_i =\frac{1}{\sqrt{N}}(s_{i1},...,s_{iN})^T, ~~~ \mbox{for $i=1,...,K$},
\end{eqnarray}
where the $s_{ia}$'s are i.i.d. random variables with zero mean and unit 
variance. These authors considered a minimum mean square error (MMSE) 
single user receiver which generates soft estimates $\hat{x}_i$'s and analyzed its performance in the large system limit 
$N,K\to\infty$, with $\alpha\equiv K/N$ constant. The analysis was 
heavily based on random matrix theory and, in particular, on the Marcenko-Pastur theorem 
on the eigenvalue distribution of large random matrices. 
As a consequence, it did not suggest any generalization to the 
practically interesting case of non-Gaussian symbols $x_i$'s.

Recently, Tanaka \cite{Tanaka} 
applied the replica method from statistical mechanics
to the case of uncoded binary antipodal signals: $x_i\in\{+1,-1\}$. 
He was able to compute the asymptotic bit
error rate and conditional entropy per bit, in the large system limit.
Given the relationship between the replica and cavity method from statistical
mechanics and message passing algorithms, such as belief propagation
(BP) and various generalizations thereof, several authors 
\cite{Kabashima,TanakaOkada} studied the
use of BP as a detection algorithm in this context.
If proved to be correct, this approach would provide a low
complexity, asymptotically optimal receiver.

Although the results obtained via replica method are likely
to be correct\footnote{Tanaka's results are derived under the
`replica symmetry' assumption. There are arguments supporting this
assumption in the problem as described so far. On the other hand, 
the same hypothesis needs to be replaced by more refined ones for 
generalizations of this problem (for instance if the noise variance
is not known at the receiver).}, the method itself is non-rigorous.
Also, the studies of BP detection algorithm were essentially empirical.
No convergence or correctness guarantees exist.

In this paper we follow the opposite route to the one sketched above.
We consider BP detection and rigorously prove convergence and correctness.
In a second step, we analyze BP to compute the system performance.
This approach solves both the problems stressed above (non rigorous
character of the replica method, and lack of guarantees on BP detection).
This paper deals with the case of Gaussian symbols $x_i$'s, 
with zero mean and unit variance. Following our strategy, 
we are able to recover several of Tse and Hanly's results without 
any recourse to random matrix theory.
We conjecture however that the same approach can be developed for
non-Gaussians symbols allowing to recover Tanaka's  results without
any recourse to the replica method.
\begin{figure}[t]
\begin{tabular}{cc}
\hspace{-1cm}
\begin{psfrags}
\psfrag{1}[]{1}
\psfrag{2}[]{2}
\psfrag{3}[]{3}
\psfrag{K}[]{$K$}
\psfrag{a}[]{$a$}
\psfrag{b}[]{$b$}
\psfrag{c}[]{$c$}
\psfrag{z}[]{$z$}
\psfrag{K Trans}[]{$K$ users}
\psfrag{N Receivers}[]{$N$ chips}
\psfrag{s1a}[]{$s_{1a}$}
\psfrag{s1b}[]{$s_{1b}$}
\psfrag{s1c}[]{$s_{1c}$}
\psfrag{s1z}[]{$s_{1z}$}
\epsfig{figure=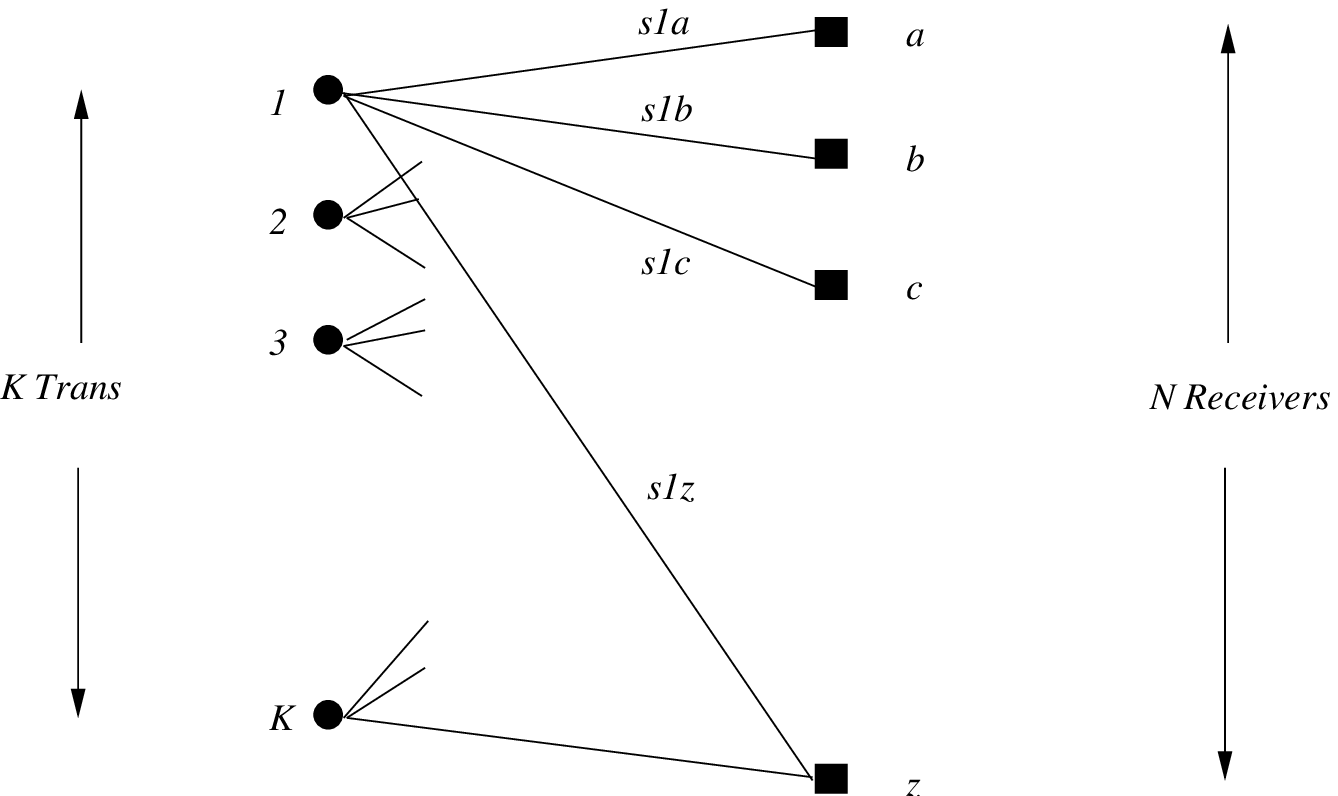, width=0.5\linewidth}
\end{psfrags}&\hspace{2cm}
\begin{psfrags}
\psfrag{1}[]{1}
\psfrag{2}[]{2}
\psfrag{3}[]{3}
\psfrag{K}[]{$K$}
\psfrag{a}[]{$a$}
\psfrag{b}[]{$b$}
\psfrag{c}[]{$c$}
\psfrag{z}[]{$z$}
\psfrag{l1a}[]{$\lambda_{1\to a}$}
\psfrag{g1a}[]{$\gamma_{1\to a}$}
\psfrag{lha1}[]{$\lh_{a\to 1}$}
\psfrag{gha1}[]{$\gh_{a\to 1}$}
\psfrag{el1}[]{$\el_1$}
\psfrag{gee1}[]{$\gee_1$}
\epsfig{figure=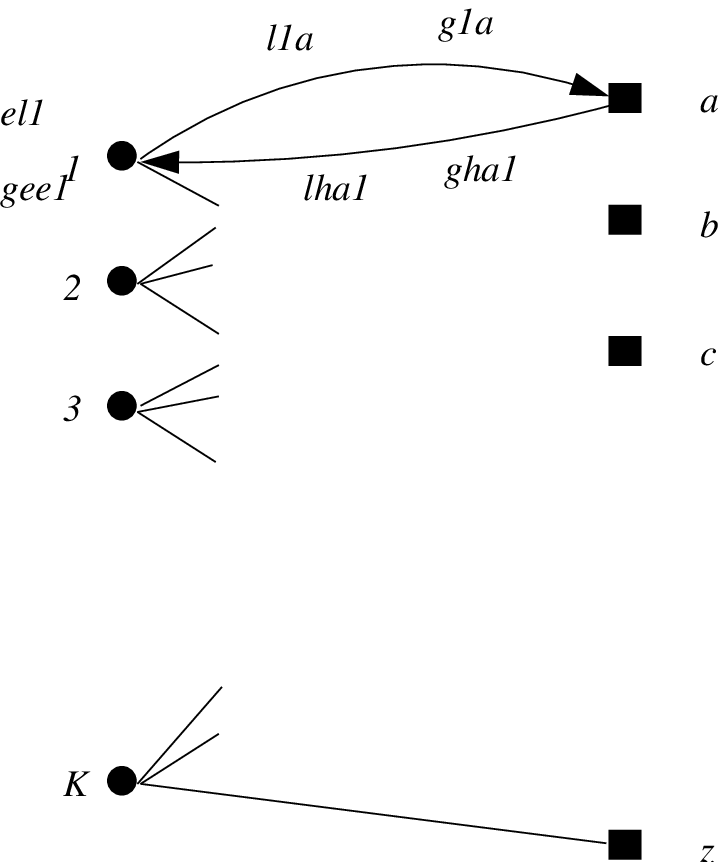, width=0.25\linewidth}
\end{psfrags}
\end{tabular}
\caption{\em Left: A graphical representation of the multiuser communication 
problem. Right: Messages notation in the BP detection algorithm. \label{fig1}}
\end{figure}

In order to apply BP, it is convenient to formulate multiuser 
detection as an inference problem on a probabilistic graphical model.
The underlying graph is depicted in Fig.~\ref{fig1}. It is a complete
bipartite graph on $K$ left nodes (users) and $N$ right nodes 
(chips). We associate real variables
$\ux\equiv (x_1,\dots,x_K)$ to the user nodes and $\uo \equiv
(\omega_1,\dots,\omega_N)$ 
to the chips, and consider the (complex) weight
\begin{eqnarray}
 \d\mu^{N,K}_y(\ux,\uo) = \frac{1}{Z^{N,K}_y}\prod_{a=1}^N
\; e^{-\frac{1}{2}\sigma^2\omega_a^2+jy_a\omega_a}
\prod_{i=1}^K \d\nu(x_i)
\prod_{i,a}
\exp\left\{- \frac{j}{\sqrt{N}}s_{ai} \omega_a x_i\right\}\,\d \omega .
\label{eq:Weight}
\end{eqnarray}
where $\d\nu(x_i) = \exp(-\frac{1}{2}x_i^2)/2\pi$ is the 
{\it a priori} distribution of the symbol $X_i$, and 
$j=\sqrt{-1}$. Elementary calculus shows
that $\d  \mu_y^{N,K}(x)  \equiv 
\int_{\omega} \d \mu_y^{N,K}(x,\omega)$ is in fact the conditional
distribution of the transmitted symbols  $x$ given $y$.
The detection problem amounts to compute the marginal
$\d  \mu_y^{N,K}(x_i) $, i.e. to integrate Eq.~(\ref{eq:Weight})
over all the variables but $x_i$. Since, the weight 
(\ref{eq:Weight}) is Gaussian, $\d  \mu_y^{N,K}(x_i) $ will be Gaussian as
well and can be parameterized as
\begin{eqnarray}
\d\mu^{N,K}_y(x_i) = \sqrt{\frac{\el_i}{2\pi}}\;
\exp\left\{-\frac{1}{2}\el_ix_i^2+\gee_i x_i\right\}\, \d x\; ,
\label{eq:MarginalParameter}
\end{eqnarray}

The weight (\ref{eq:Weight}) factorizes according to the complete
bipartite graph in Fig.~\ref{fig1}, left frame. BP can therefore 
be used to computing the marginals $\d  \mu_y^{N,K}(x_i) $
through a message passing procedure. Messages are exchanged as
in Fig.~\ref{fig1}, right frame, and computed according to the
update equations
\begin{eqnarray}
\lambda^{(t+1)}_{i\to a} = 1+\frac{1}{N}\sum_{b\neq a}\frac{s_{ib}^2}
{\lh^{(t)}_{b\to i}}\, ,
\;\;\;\;\;\;\;\;\;\;\;
\lh^{(t)}_{a\to i} = \sigma^2+\frac{1}{N}\sum_{k\neq i}
\frac{s_{ka}^2}{\lambda^{(t)}_{k\to a}}\, \label{eq:VariancesBPGaussian}  \\
\gamma^{(t+1)}_{i\to a} = \frac{1}{\sqrt{N}}\sum_{b\neq a}
\frac{s_{ib}}{\lh^{(t)}_{b\to i}}\, \gh^{(t)}_{b\to i}\, ,\;\;\;\;\;\;\;
\gh^{(t)}_{a\to i} = y_a -\frac{1}{\sqrt{N}}\sum_{k\neq i}
\frac{s_{ka}}{\lambda^{(t)}_{k\to a}}\, \gamma^{(t)}_{k\to a}\, .
\label{eq:MeansBPGaussian}
\end{eqnarray}
with initial conditions $\lambda^{(0)}_{i\to a} = 1$,
$\gamma^{(0)}_{i\to a} = 0$.
The parameters in the marginal distribution $\d\mu^{N,K}_y(x_i)$,
cf. Eq.~(\ref{eq:MarginalParameter}), are estimated as
\begin{eqnarray}
\gee_i^{(t+1)} = \frac{1}{\sqrt{N}}\sum_{b\in [N]}
\frac{s_{ib}}{\lh^{(t)}_{b\to i}}\, \gh^{(t)}_{b\to i}\, ,\;\;\;\mbox{and}\;\;\;\;
\el_i^{(t+1)} = 1+\frac{1}{N}\sum_{b\in [N]}\frac{s_{ib}^2}
{\lh^{(t)}_{b\to i}}\, .
\end{eqnarray}

In the next Section we will show that this procedure is convergent
and provides the asymptotically correct marginal distributions,
thus implementing allowing to implement a MMSE estimator.
In Section \ref{sec:Recover} we show that the usual prescription for
a MMSE receiver can be recovered from the fixed point of BP.
Finally, in Section \ref{sec:Numerical}, we investigate the rate of
convergence of our algorithm through numerical simulations.
%
%
\section{Main result: convergence and correctness}

Throughout this Section we  assume that $s_{ia} = \pm 1$ with equal 
probability. We believe that our main results, as well as the idea of
the proof, remain valid for a considerably more general distribution.
However, this assumption allows to avoid several technical complications.
In particular, it is immediate to prove the following result
\begin{lemma}\label{VarianceLemma}
If $s_{ia}\in\{+1,-1\}$ for any $i\in [K]$, $a\in[N]$, then 
$\lambda_{i\to a}^{(t)} = \lambda(t)$, $\lh_{a\to i}^{(t)} = \lh(t)$,
where $\lambda(t), \lh(t)$ are given by the iterations
\begin{eqnarray}
\lambda(t+1) = 1+\frac{N-1}{N}\; \frac{1}{\lh(t)}\, ,\;\;\;\;\;\;\;
\lh(t) = \sigma^2+\frac{K-1}{N}\;\frac{1}{\lambda(t)}\, ,
\end{eqnarray}
with $\lambda(0) = 1$.

Moreover $\lambda(t)\to\lambda^{(\infty)}$, $\lh(t)\to\lh^{(\infty)}$ 
as $t\to\infty$,
with $\lambda^{(\infty)},\lh^{(\infty)}>0$ the unique positive fixed point of the above 
equations.  More precisely, we have $\lambda^{(\infty)} = 1+\Lambda$,
$\lh^{(\infty)} = 1/\Lambda$, where
\begin{eqnarray}
\frac{1}{\Lambda} = \sigma^2+\frac{\alpha}{1+\Lambda}\, .\label{eq:David}
\end{eqnarray}
\end{lemma}
The reader will recognize that Eq.~(\ref{eq:David}) is nothing but the
Tse-Hanly asymptotic formula for the signal to interference ratio in the
large system limit.

We shall state two separate results for the mean of the marginal
distribution (\ref{eq:MarginalParameter}), 
$\ex_i = \gee_i/\el_i$ and its variance 
$1/\el_i$. 
\begin{thm}
\label{name}
Assume the $s_{ia}$'s are i.i.d. uniformly random in $\{+1,-1\}$.
Then, there exists a set ${\cal S}_N\subseteq \{+1,-1\}^{K\times N}$ of 
signatures with $\prob({\cal S}_N)\ge 1-O(N^{-a})$ (with $a$ an arbitrary
positive number) 
such that, for any  ${\mathbb S}\in {\cal S}_N$ and for every $\{y_a\}$,
$\el_i^{(t)}\to\el_i^{(\infty)}$ and 
$\gee_i^{(t)}\to\gee_i^{(\infty)}$ as $t\to\infty$.  
Furthermore, the conditional means estimated by Belief Propagation 
are correct,
i.e., $\ex^{(\infty)}_i  = \gee_i^{(\infty)}/\el_i^{(\infty)} = \ex_i$.
\end{thm}
{\bf Sketch of proof:} For Gaussian graphical models, it was proved in 
\cite{FreemanWeiss,VanRoy} that, if BP converges for a `generic initial 
condition', then it correctly estimates the mean of the associated 
probability distribution. Referring to the present case, this immediately
implies $\ex^{(\infty)}_i = \ex_i$ if 
convergence  holds for any initial condition of the form 
$\gamma^{(0)}_{i\to a} = c_{i\to a}$, $\lambda^{(0)}_{i\to a} = 1$.

Because of Lemma~\ref{VarianceLemma}, it is in fact sufficient to
prove convergence for the messages $\{\gamma^{(t)}_{i\to a}\}$.
We consider therefore the homogeneous recursion
\begin{eqnarray}
\Gamma^{(t+1)}_{i\to a} = \frac{1}{\sqrt{N}\lh(t)}\sum_{b\neq a}
s_{ib} \Gh^{(t)}_{b\to i}\, ,\;\;\;\;\;\;\;
\Gh^{(t)}_{a\to i} =  -\; \frac{1}{\sqrt{N} \lambda(t)}\sum_{k\neq i}
s_{ka}\, \Gamma^{(t)}_{k\to a}\, .
\end{eqnarray}
We shall show that for a typical realization of the signatures $\si$,
$\Gamma^{(t)}_{i\to a}$ and $\Gh^{(t)}_{a\to i}\to 0$ for any initial 
condition $\Gamma^{(0)}_{i\to a}$ and $ \Gh^{(0)}_{a\to i}$.  This in turns
imply that $\gamma^{(t)}_{i\to a}$ and $\gh^{(t)}_{a\to i}$ converge.
We begin by eliminating $\Gh^{(t)}_{a\to i}$ from the above relations to
get
\begin{eqnarray}
\Gamma^{(t+1)}_{i\to a} = -\; \frac{1}{N\lambda(t)\lh(t)}
\sum_{k\to b }\Omega_{ia,kb} \Gamma^{(t)}_{k\to b}
\end{eqnarray}
where 
\begin{eqnarray}
\Omega_{ia,kb} = \left\{\begin{array}{ll}
0 & \mbox{   if $i=k$ or $a=b$,}\\
s_{ib}s_{kb}& \mbox{   otherwise.}
\end{array}\right. \label{eq:OmegaDef}
\end{eqnarray}
The random matrix $\Omega = \{\Omega_{ia,kb} : i,k\in [K], a,b\in [N]\}$
has dimensions $NK\times NK$ and is a function of the signature sequence
$\si$.

We make use of the following
property of $\Omega$, whose proof is omitted.
\begin{lemma}\label{TraceLemma}
For each $\alpha>0$, and each $\kappa>1$ there exist a positive integer
$N_0=N_0(\alpha,\kappa)$ such that, if $N>N_0$ and $t\le (N/1000)^{1/6}$
then
\begin{eqnarray}
\E\,\Tr\left\{(\Omega^t)^{\dagger}\Omega^t\right\} \le N^{2t+2}
\alpha^{t+1}\,\left(1+ \frac{C(\alpha,\kappa)}{N}\, t^6\, \kappa^t\right)\, 
,\label{Trace}
\end{eqnarray}
where $C(\alpha,\kappa)$ is $N$-- and $t$--independent. 
\end{lemma}

Let $\zeta_{\rm max}$ be the eigenvalue of $\Omega$ with the largest 
absolute value, and denote by $\cS_N(\rho)\subseteq \{+1,-1\}^{NK}$
the set of signature sequences, such that 
$|\zeta_{\rm max}|< N\rho\lambda^{(\infty)}\lh^{(\infty)}$. 
Applying Markov
inequality to the random variable 
$|\zeta_{\rm max}|^{2t}\le \Tr\left\{(\Omega^t)^{\dagger}\Omega^t\right\}$,
it is easy to show that
\begin{eqnarray}
\prob\left\{\overline{\cS_N(\rho)}\right\}&\le & 
N^{2}\alpha \left(\frac{\sqrt{\alpha}}{\rho\lambda^{(\infty)}
\lh^{(\infty)} }\right)^{2t}
\left(1+ \frac{C(\alpha,\kappa)}{N}\, t^6\, \kappa^t\right)\, .
\end{eqnarray}
By properly choosing  $\rho$ and $t$ (and setting $\cS_N = \cS_N(\rho)$), 
this in turns imply $\prob\left\{\overline{\cS_N}\right\}\le O(N^{-a})$
with $a>0$.

We must now prove that, for any signature sequence in
$\cS_N$, 
$\Gamma^{(t)}\equiv [\Gamma^{(t)}_{i\to a}]\to 0$.
Since $\lambda(s)\to\lambda^{(\infty)}$, $\lh(s)\to\lh^{(\infty)}$,
there exists $s^*$, such that $|\zeta_{\rm max}|<N\rho\lambda(s^*)\lh(s^*)$,
with $\rho<1$.
Defining $\widehat{\Omega} = - \Omega/N\lambda(s^*)\lh(s^*)$, we have
\begin{eqnarray}
\Gamma^{(t)}  = \prod_{s=0}^{t-1}
\left[\frac{-1}{N\lambda(s)\lh(s)}\,
\Omega\right] \, \Gamma^{(0)} 
= \left\{\prod_{s=s^*}^{t-1}
\frac{\lambda(s^*)\lh(s^*)}{\lambda(s)\lh(s)}\right\}
\widehat{\Omega}^{t-s^*} \! \Gamma^{(s_*)}
\, .
\end{eqnarray}
Since the largest eigenvalue of  $\widehat{\Omega}$ has modulus
smaller than one,   $\widehat{\Omega}^{t-s^*}\! \Gamma^{(s_*)}\to 0$.
Furthermore, since $\lambda(s)$, $\lh(s)$ increases with $s$, 
the number in curly brackets is smaller than one. Therefore
$\Gamma^{(t)}\to 0$ as well.\hspace{9.cm} $\Box$

Unlike for the means $\ex_i$, the BP estimate of variances
is only correct in the large system limit $N,K\to \infty$.
In order to quantify the error made, we define 
$\delta_i=(1/L^{(\infty)}_i)-(1/L_i)$ and
\begin{eqnarray}
D\equiv \frac{1}{K}\sum_{i=1}^K \delta_i^2\, .
\end{eqnarray}
\begin{thm}\label{VarianceThm}
Assume the $s_{ia}$'s are i.i.d. uniformly random in $\{+1,-1\}$,
then $D\to 0$ in probability as $N\to\infty$.
\end{thm}
{\bf Sketch of proof:} Freeman and Weiss \cite{FreemanWeiss}
provide an explicit expression for the discrepancies $\delta_i$.
In the present case, their expression can be shown to be equivalent
to the following one
\begin{eqnarray}
\delta_i = \frac{1}{L^{(\infty)}}\sum_{\pi:i\to i}
\left(\frac{-1}{N\lambda^{(\infty)}\lh^{(\infty)}}\right)^{|\pi|/2}
\prod_{(k,b)\in \pi}s_{kb} \, .\label{eq:DeltaPath}
\end{eqnarray}
Here the sum runs over all the closed non-reversing paths $\pi$,
on the bipartite graph in Fig.~\ref{fig1},
starting and ending at $i$. 

In order to convey the basic idea of the proof, let us restrict the above 
sum to the paths of a given {\em fixed} length $2t$, and call
$\delta_i^{(t)}$ the corresponding quantity. It is easy to show that
\begin{eqnarray}
\E(\delta_i^{(t)})^2 = \frac{1}{(L^{(\infty)})^2}
\left(\frac{1}{N\lambda^{(\infty)}\lh^{(\infty)}}\right)^{2t}
\left|\J^i_{N,K}(2t)\right|\, ,
\end{eqnarray}
where $\J^i_{N,K}(2t)$ is the set of {\em couples} of closed paths 
starting at $i$ on the bipartite graph in Fig.~\ref{fig1}, 
such that each edge in the graph is visited an {\em even} number of times.
The dominating contribution to $|\J^i_{N,K}(2t)|$ comes, in the
large system limit, from couples of coincident paths.
Their number scales like $N^{2t-1}$: we are free to choose each step but the
last one among $O(N)$ vertices. Therefore 
$\E(\delta_i^{(t)})^2 = O(N^{-1})\to 0$. 

In order to complete the
proof, one has to control those paths in Eq.~(\ref{eq:DeltaPath}),
whose length diverges with $N$.\hspace{10.7cm}
$\Box$

%
%
\section{MMSE receiver recovered}
\label{sec:Recover}

The per-iteration complexity of the BP algorithm considered above,
scales $N^3$. In fact the number of messages is equal to the number of 
edges in the complete bipartite graph of 
Fig.~\ref{fig1}, i.e. $O(N^2)$, and each message update involves $O(N)$ 
sums.  
This may be too much for some applications. 
Also, although we proved that BP returns the correct MMSE estimate
$\ex$, the relation between the two approaches is quite puzzling.
Recall that MMSE is given by the explicit formula
\begin{eqnarray}
\ex(y) = [\sigma^2\ind +\si^\dagger \si]^{-1}\si^{\dagger}y\, .
\label{eq:ExGeneral}
\end{eqnarray} 
where $\si$ is the $N$ by $K$ matrix whose columns are the signature sequences of the users.
This formula only involve vectors of size $O(N)$. BP, on the other
hand, involved linear operations on the vector $[\gamma_{i\to a}^{(t)}]$,
of size $NK$.

In  order to clarify the relationship between the two approaches,
let us write
\begin{eqnarray}
\gamma_{i\to a}^{(t)} = \gee^{(t)}_i+\dg^{(t)}_{i\to a}\, ,\;\;\;\;\;\;\;
\gh_{a\to i}^{(t)} = \geeh^{(t)}_a+\dgh^{(t)}_{a\to i}\, .
\end{eqnarray}
The BP update equations  for $\gamma^{(t)}_{i\to a}$,
cf.~Eq.~(\ref{eq:MeansBPGaussian}), read in the new
variables (to lighten the formulae we assume here $s_{ia}\in \{+1,-1\}$)
\begin{eqnarray}
\gee^{(t+1)}_i &=& \frac{1}{\lh(t)\sqrt{N}}\sum_{b=1}^N
s_{ib}\geeh^{(t)}_b+\frac{1}{\lh(t)\sqrt{N}}
\sum_{b=1}^N s_{ib}\dgh^{(t)}_{b\to i}\, ,\label{Vertex}\\
\dg^{(t+1)}_{i\to a} & = & -\frac{1}{\lh(t)\sqrt{N}}s_{ia}\geeh^{(t)}_b
-\frac{1}{\lh(t)\sqrt{N}} s_{ia}\dgh^{(t)}_{a\to i}\, .\label{Delta}
\end{eqnarray}
Analogously, the update equations for $\gh^{(t)}_{a\to i}$ become
\begin{eqnarray}
\geeh^{(t)}_a &=& y_a-\frac{1}{\lambda(t)\sqrt{N}}\sum_{k=1}^K
s_{ka}\gee^{(t)}_k-\frac{1}{\lambda(t)\sqrt{N}}
\sum_{k=1}^K s_{ka}\dg^{(t)}_{k\to a}\, ,\label{Vertexh}\\
\dgh^{(t)}_{a\to i} & = & \frac{1}{\lambda(t)\sqrt{N}}s_{ia}\gee^{(t)}_i
+\frac{1}{\lambda(t)\sqrt{N}} s_{ia}\dgh^{(t)}_{i\to a}\, .\label{Deltah}
\end{eqnarray}
From Eqs.~(\ref{Delta}), (\ref{Deltah}), it follows that
$\dg^{(t)}_{i\to a},\dg^{(t)}_{i\to a} = O(N^{-1/2})$ and are 
given approximately by
\begin{eqnarray}
\dg_{i\to a}^{(t)} = -\frac{1}{\lh(t-1)\sqrt{N}}s_{ia}\geeh_a^{(t-1)}
+O(N^{-1})\,
,\;\;\;\;
\dgh_{a\to i}^{(t)} = \frac{1}{\lambda(t)\sqrt{N}}s_{ia}\gee_i^{(t)}+
O(N^{-1})\, .\!\!\!\!\!\!\!\!\!\!\nonumber\\
\end{eqnarray}
Substituting in Eqs.~(\ref{Vertex}), (\ref{Vertexh}), and neglecting 
$O(N^{-1})$ terms, we get, with a slight abuse of notation
\begin{eqnarray}
\gee^{(t+1)}_i &=& \frac{1}{\lambda(t)\lh(t)}\,\gee_i^{(t)}
+\frac{1}{\lh(t)\sqrt{N}}\sum_{b=1}^N s_{ib}\geeh^{(t)}_b\, ,
\label{eq:ApproxBpEq1}\\
\geeh^{(t)}_a &=& y_a+\frac{\alpha}{\lambda(t)\lh(t-1)}\,\geeh_a^{(t-1)}
-\frac{1}{\lambda(t)\sqrt{N}}\sum_{k=1}^K s_{ka}\gee^{(t)}_k\, .
\label{eq:ApproxBpEq2}
\end{eqnarray}
These update equations will be referred to as `approximate belief propagation'.
As opposed to ordinary BP, they involve quantities associated to
the vertices of the underlying graphical model. Their complexity
per iteration is $O(N^2)$, which can be compared to the $O(N^3)$
complexity of ordinary BP for the present problem.

Let us look for a fixed point $\gee^{(\infty)} = [\gee_i^{(\infty)}]$,
$\geeh^{(\infty)} =[\geeh^{(\infty)}_a]$ 
of Eqs.~(\ref{eq:ApproxBpEq1}), (\ref{eq:ApproxBpEq2}) 
and neglect $O(N^{-1})$ terms
for the time being. Recall that, at the fixed point
$\lambda^{(\infty)} = 1+\Lambda$, and $\lh^{(\infty)} = 1/\Lambda$,
where $\Lambda$ solves Eq.~(\ref{eq:David}).
Substituting in Eqs.~(\ref{eq:ApproxBpEq1}), (\ref{eq:ApproxBpEq2}),
and adopting matrix notation we obtain the fixed point conditions
\begin{eqnarray}
\gee^{(\infty)} & = & \frac{\Lambda}{1+\Lambda}\, \gee^{(\infty)}+
\Lambda \si^{\dagger}\geeh^{(\infty)}\, ,\label{eq:ApproxFP1}\\
\geeh^{(\infty)} & = & \frac{\alpha\Lambda}{1+\Lambda}\, \gee^{(\infty)}
+y-\frac{1}{1+\Lambda}\si\, \gee^{(\infty)}\, .\label{eq:ApproxFP2}
\end{eqnarray}
As explained in Sec.~\ref{sec:Intro}, the conditional expectations 
estimates are given by $\ex^{(\infty)} = \gee^{(\infty)}/\el^{(\infty)}$.
It is not hard to realize that $\el^{(\infty)} = \lambda^{(\infty)}+O(N^{-1})=
1+\Lambda+O(N^{-1})$. By solving Eqs.~(\ref{eq:ApproxFP1}) and (\ref{eq:ApproxFP2})  for 
$\gee^{(\infty)}$ and using this formula, one finally gets
$\ex^{(\infty)} = \ex(y)$, where $\ex(y)$ is given by
Eq.~(\ref{eq:ExGeneral}).

The above discussion has important practical implications. One can
replace the original BP equations, with 
Eqs.~(\ref{eq:ApproxBpEq1}), (\ref{eq:ApproxBpEq2}), which have lower 
complexity. This provides an `approximate BP' algorithm.
A more careful examination of the above steps allows to prove the 
following statement.
\begin{propo}
Assume the $s_{ia}$'s are i.i.d uniformly random in $\{+1,-1\}$.
Then, with high probability, approximate BP is convergent,
$\gee^{(t)}\to\gee^{(\infty)}$, $\el^{(t)}\to\el^{(\infty)}$, and 
provides correct estimates of the conditional means,
i.e. $\ex^{(\infty)}\equiv \gee^{(\infty)}/\el^{(\infty)} = \ex(y)$.
\end{propo}
%
%
\section{Numerical simulations and convergence rate}
\label{sec:Numerical}

In the previous Section we discussed the complexity per iteration of the 
BP detection algorithm. We found that it can be as small as $O(N^2)$
if the `approximate' update equations (\ref{eq:ApproxBpEq1}), 
(\ref{eq:ApproxBpEq2}) are used. Optimal (MMSE) detection is achieved
after convergence to the BP fixed point.

One may wonder how many iterations are required for the algorithm to 
get `reasonably close' to its fixed point. A first answer is
provided by the proof of Theorem \ref{name}. Each BP iteration 
involves the multiplication of the vector of messages by the matrix 
$\Omega$, and the division by $N\lambda(t)\lh(t)$. Lemma 
\ref{TraceLemma} suggests in turn that the eigenvalues of 
$\Omega$ have absolute value not larger than $N\sqrt{\alpha}$. Therefore,
at each iteration, the distance between current vector of messages and 
the the fixed point messages is, roughly speaking,
rescaled by a factor 
\begin{eqnarray}
\frac{\sqrt{\alpha}}{\lambda(t)\lh(t)}\approx \frac{\sqrt{\alpha}\, \Lambda}{
1+\Lambda}\, ,
\end{eqnarray}
where we approximate the quantities on the left hand side by their value 
at the fixed point.
\begin{figure}[t]
\begin{tabular}{cc}
\hspace{-1cm}
\epsfig{figure=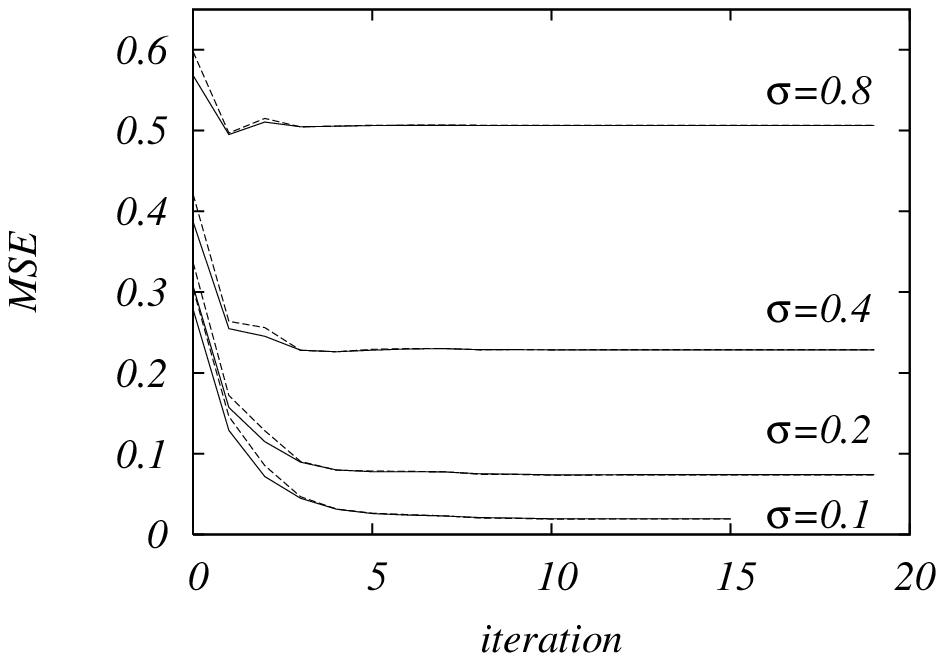, width=0.45\linewidth}&
\epsfig{figure=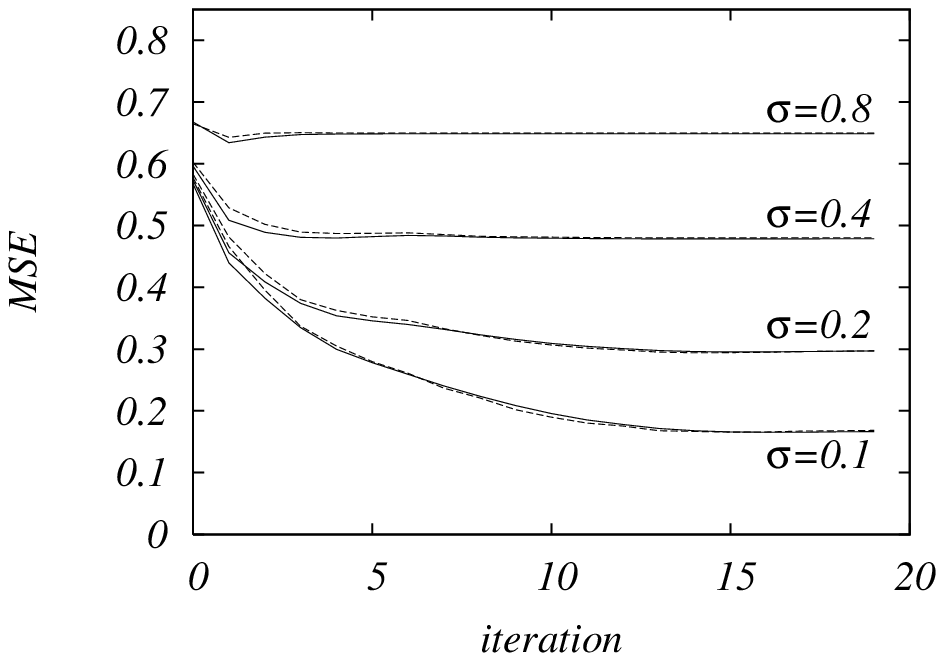, width=0.45\linewidth}
\end{tabular}
\caption{\em Mean square error for BP based multiuser detection, as a function
of the number of iterations. The asymptotic values coincide with the
ones for MMSE detection. Continuous lines refer to ordinary BP, 
and dashed lines to approximate BP. Left: $K=100$ users and 
$N=200$ chips ($\alpha = 0.5$). Right: $K=100$ users and 
$N=100$ chips ($\alpha = 1$). \label{fig2}}
\end{figure}
An equivalent way of formulating this result consists in defining
\begin{eqnarray}
t_*(\sigma,\alpha)  = -\left(\log \frac{\sqrt{\alpha}\Lambda}{1+\Lambda}
\right)^{-1}\, ,\label{eq:Tstar}
\end{eqnarray}
where $\Lambda$ is understood to be the solution of Eq.~(\ref{eq:David}).
After $t_*= t_*(\sigma,\alpha)$ iterations, the distance from the fixed
point is rescaled by a constant factor $e^{-1}$. Any precision
$\delta$ can be achieved within a number of iterations 
$t(\delta) \approx t_*(\sigma,\alpha)\, \log(\Delta/\delta)$.
where $\Delta$ is the distance between the initial and 
fixed point messages. 
 
It can be shown that, 
the BP estimate after the first iteration is equal to the one provided by the 
traditional matched filter receiver. As the number of iterations
increases, BP moves from the matched filter to the MMSE estimate. 
After about $t_*(\sigma,\alpha)\log(1/\varepsilon)$ iterations,
the distance from the MMSE estimate is reduced by a factor $\varepsilon$
with respect to a matched filter receiver.

In order to confirm this analysis and obtain some 
concrete feeling of the actual number of required BP iterations,
we plot in Fig.~\ref{fig2} the results of  numerical simulations 
of the BP based detection algorithm. Each couple of curves refers to the application
of BP (continuous curves) or approximate BP (dashed curves) to a single 
signature/noise realization. Convergence to the MMSE receiver
was achieved in each case. 

A small number of iterations (about $5$)  provides a large improvement 
over the matched filter receiver. The number of iterations
required for convergence is always moderate and consistent 
with the estimate of Eq.~(\ref{eq:Tstar}).
For $\alpha = 0.5$ (left frame), we obtain, for instance, 
$t_*\approx 2.7$, $2.4$, $1.7$ and $1.0$ for (respectively) 
$\sigma = 0.1$, $0.2$, $0.4$ and $0.8$.
If $\alpha = 1$ (right frame), we obtain  
$t_*\approx 10.0$, $5.0$, $2.5$ and $1.3$ for the same values of $\sigma$.
%
%
\section{Conclusion and generalizations}

We can envision two types of generalizations of the present work.
First, one can refine the model (\ref{eq:Model}) by considering, for
instance, different power constraints for different users, or
signature sequences with more general distributions than the one considered 
in this paper. While such generalizations can be technically cumbersome,
they should possible along the same lines exposed here.

The second direction consists in modifying the {\it a priori} distribution
of the signals $X_i$, by considering, for instance, 
binary antipodal signals, $x_i\in\{+1,-1\}$.
The proves of  Theorems \ref{name} and \ref{VarianceThm} are heavily based
on the general results on belief propagation for Gaussian graphical models
in \cite{FreemanWeiss} and \cite{VanRoy}. These results do not
extend to symbols $x_i$'s with a general {\it a priori}
distribution. However, we expect the thesis (convergence and correctness
of belief propagation) to remain true in the large system limit.
We plan to use the ideas of \cite{itw} to prove this claim.
%
%
\section*{Acknowledgments}

AM has been partially supported by EVERGROW, i.p. 1935 of the
EU Sixth Framework.

\end{document}